# Visuohaptic augmented feedback for enhancing motor skills acquisition

Ali Asadipour[1] · Kurt Debattista[1] · Alan Chalmers[1]

**Abstract** Serious games are accepted as an effective approach to deliver augmented feedback in motor (re-)learning processes. The multi-modal nature of the conventional computer games (e.g. audiovisual representation) plus the ability to interact via haptic-enabled inputs provides a more immersive experience. Thus, particular disciplines such as medical education in which frequent hands on rehearsals play a key role in learning core motor skills (e.g. physical palpations) may benefit from this technique. Challenges such as the impracticality of verbalising palpation experience by tutors and ethical considerations may prevent the medical students from correctly learning core palpation skills. This work presents a new data glove, built from off-the-shelf components which captures pressure sensitivity designed to provide feedback for palpation tasks. In this work the data glove is used to control a serious game adapted from the infinite runner genre to improve motor skill acquisition. A comparative evaluation on usability and effectiveness of the method using multimodal visualisations, as part of a larger study to enhance pressure sensitivity, is presented. Thirty participants divided into a game-playing group ($n = 15$) and a control group ($n = 15$) were invited to perform a simple palpation task. The game-playing group significantly outperformed the control group in which abstract visualisation of force was provided to the users in a blind-folded transfer test. The game-based training approach was positively described by the game-playing group as enjoyable and engaging.

## 1 Introduction

The efficacy of serious games as a training approach has become widely accepted and as a consequence serious games are beginning to be used in a wide variety of domains. Immersion, pleasure and competition are key characteristics of games that enhance user engagement in training activities [1]. One of the main domains demonstrating the fruitfulness of serious games is healthcare [2]. A number of reasons have led to the success of serious games for healthcare. Ethical restrictions in medical training mean that certain procedures cannot be frequently tested and explored by practitioners. Similarly, a lack of available patients during training entails limited familiarity with the application of procedures across a varying number of ages, genders and body types. Furthermore, the demands for patients' growing thirst for health information has led to health professionals providing novel digital-based interventions. Serious games are used in this work as a means of providing augmented visuohaptic feedback that enables users to synchronise their visual and kinaesthetic perceptions. Game-based learning provides effective motivation via enjoyable experiences and enhanced user engagement.

Current challenges in medical education are particularly difficult when the possibilities of experiencing training in that activity are rare. Game-based solutions can provide an enhanced experience to the existing educational process for such cases [3,4]. One of the most common medical processes which are carried out by most medical practitioners in a very large variety of conditions and for a diverse number of appli-cations is palpation [5,6]. Palpation plays an important role in the initial examination of patients and is a crucial initial diagnosis [7] based significantly on haptic sensory feedback. While visual and acoustic digital healthcare practices, and audio-visual cross-modal research [8], are becoming more common due to the ubiquitous nature of video and audio displays and their use in both entertainment and real-world applications, the lack of readily available haptic devices has meant very little progress has been made in the areas of motion and pressure sensitivity control within the healthcare domains.

✉ Ali Asadipour
A.Asadipour@warwick.ac.uk

[1] Visualisation Group, WMG, University of Warwick, Coventry, UK

This work is part of a larger project aimed at training medical students to become more proficient at palpation as part of their training process [9]. Training in palpation is of crucial importance as during their training medical students are restricted to the number of hours they can spend with their tutors gaining hands-on experience and are also restricted on the number of body types and participants that they can engage with [10]. A goal of automating the process and providing palpation-based simulators will enhance current practices by allowing the students to practice by themselves or with each other while being guided via a digital tutor. This paper identifies and focuses on one aspect of an automated palpation framework: training of pressure sensitivity using the index finger, one of the crucial characteristics of palpation training [11–13], is provided via the use of a serious game in which the player controls an on-screen character via the use of an input device which is sensitive to pressure. Learning to apply the correct amount of pressure plays a significant role in providing the correct diagnosis and also in patient comfort; a too light touch may miss out on important physiological phenomena and a too heavy touch may cause significant patient discomfort further compounding potential diagnosis issues.

While a number of novel input technologies beyond the traditional have recently begun to be applied to serious games [14–17], no serious game, to the best of our knowledge, has targeted the correct application of pressure as its main goal. This work introduces the ParsGlove, an input device used for measure pressure sensitivity built using off-the-shelf components; details on how it is calibrated and engineered are provided. A study based on two groups composed of general public participants, one group that played the game and a control group demonstrates that there is reason to believe that such a serious game, can help improve pressure sensitivity in individuals.

The following section presents background and related work. Descriptive information of the three technologies which are used in this study is discussed in Sect. 3. The experimental design and results are discussed in Sect. 4. Finally, future potentials for extending this study and conclusions are presented in Sect. 5.

## 2 Background and related work

In general, applications of serious games in healthcare can be classified by their target audience [18] as follows:

- Medical education
- Patient intervention
- Public involvement

Games dedicated to medical education are those that help medical professionals to improve their skills while performing certain tasks. Patient intervention is targeted at patients rather than the medical professionals. Such patient-oriented training games help enhance individuals' knowledge about their condition and to also help improve their engagement in their treatment process. Public involvement applications are directed at the general public and are focused on raising awareness of public health issues and providing motivation for potential behavioural change. A large number of healthcare-related serious games have been developed [2] and we provide a small set of examples in the following:

In terms of medical education and awareness Graafland et al. [19] presented a survey on medical education and surgical skills across 25 publications which comprised 30 games. They explained that games developed for the purpose of such serious applications required the use of further validation before being deployed as there was a lack of robust evaluation for the surveyed games. Dunwell et al. [20] presented a serious game to help create awareness of healthcare-associated infections within wards. They provide feedback and findings on the deployment of the game across 13 hospital wards in the United Kingdom. The serious game we present in this paper could also be considered as part of the medical education sub-category of serious games.

An example of a patient intervention game was the Re-Mission game for cancer patients. Positive behavioural changes were reported on pediatric patients who were diagnosed with cancer by playing this video game [21]. Another patient intervention example was provided by Carmeli et al. [22]. The authors presented a serious game for the improvement of motor, sensory and cognitive performance in rehabilitation of stroke patients (with upper limb impairments) to conduct everyday functional tasks efficiently. An interactive tool was used as an input device for the game to measure range of motion and finger and wrist speed. Results demonstrated an improvement in movements for users.

An example of increasing public involvement was presented by Boulos et al. [23] environment to raise public awareness about sexual health. An online 3D virtual world such as Second Life (SL) with social networking capabil-ities has been surveyed to highlight positive impact on its audience. Brown et al. [24] also presented a serious game dealing with sex education in which an intervention mapping approach was used in the development and the design of the game.

**Fig. 1** ParsGlove is an innovative wearable interface which is designed and developed to capture the human hand ergonomics and to provide this information as an interactive input both for the game and the application

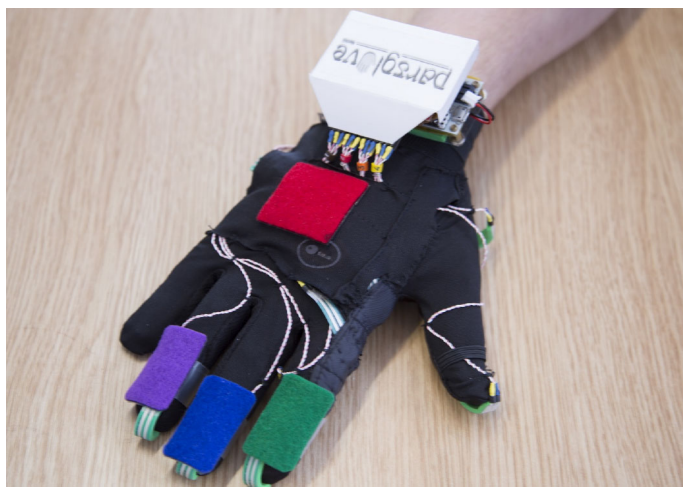

Scarle et al. [15] presented a serious game that had two main goals focusing on public involvement. First, it was targeted at raising the awareness of poor eating habits at primary school children that has been becoming one of the main causes of the rising obesity epidemic and, second, through the application of motion controls that enabled the participant to reduce the amount of on-screen time in which the game player was physically inactive.

Srinivasan and Chen [25] have investigated the human ability in controlling normal forces and the impact of various augmented sensory feedback in aiding control performance. Three human subjects were asked to exert forces with their index fingertip on a force sensor while seated in a force tracking experiment. The target force–time profile was displayed at the beginning of each trial and participants were instructed to track the target force as close as possible. Local anaesthesia was administered to the middle phalanx of the subject's index finger in a follow up experiment, to block the tactile sensory information; in the first experiment under normal condition both tactile and kinaesthetic information were available to the subjects. In constant force tracking tasks, visual feedback was provided under both normal and anaesthetised conditions and it was later withdrawn to repeat the tasks. Absolute error (error = |desired − actual|) was computed for each constant force targets (0.2 to 1.5 N in 0.25 N steps). Results showed that the magnitude of the error and its variations were affected by the absence of visual feedback (under both normal and anaesthetised conditions) with respect to target force values whereas only the magnitude of the error was affected by the absence of the tactile sensation.

## 3 Serious game and input device

This section presents the overall framework that has been used for this work. Three key technologies including a serious game were developed to improve pressure sensitivity learning. The input device is a glove developed in-house and can read pressure the amount of pressure applied accurately in Newtons. An application, DigiScale, was developed to help facilitate the input procedure via a user-friendly interface. Finally, a serious game was implemented to facilitate the learning of pressure sensitivity for the user.

### 3.1 Input device

A wearable measurement interface, ParsGlove (see Fig. 1) has been developed under formal research and development discussions with medical professionals. It is designed to capture the ergonomics of the human hand during dexterous interaction with the environment. It was essential in our main goals to use the full capabilities of the glove to capture applied pressures, orientations and location parameters although for this work the focus is only on the pressure input.

To provide freedom for the practitioners, the glove is equipped with Bluetooth connectivity and is composed of ultralight materials which help reduce weight and avoid fatigue when the glove is worn for long periods. Twelve force-sensitive resistors are mounted in places which were defined by medical professionals. Sensors are manufactured by IEE Ltd. [26] with 2–3 ms typical response time on single activation and $2 \pm 0.2$ mm thickness of pad. More technical details are presented in the next section. The sensors were also calibrated with a force gauge device to accurately map digital values to actual force as described below.

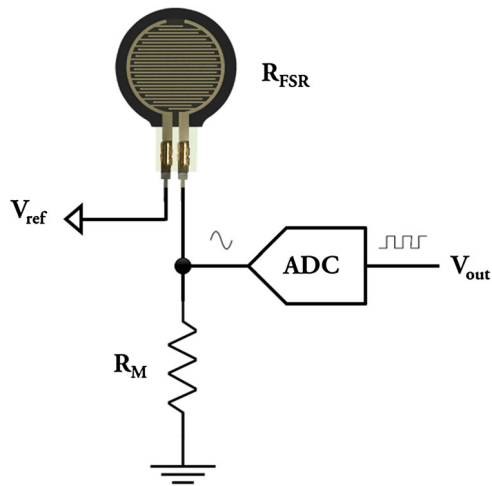

**Fig. 2** FSR sensors' suggested circuitry schematic

### 3.2 Calibration

Piezoresistive force-sensitive resistors (FSR) are widely used in ergonomics studies [27,28,28] to measure magnitude of applied forces by different parts of the human body. The low-cost per unit, simplicity in calibration, ease of use, their availability in different shapes and sizes and close to linear behaviour in lower force ranges are some of the advantages of the FSR sensors. FSR sensors' measurement range are highly affected by their circuitry interfacing method. Figure 2 shows the schematic suggested by the manufacturer for interfacing these sensors.

The electrical resistance ($R_{FSR}$) of the conductive material which is used in these type of sensors decreases by increasing the applied forces on the sensor's active area. This resistance is an infinite value in the absence of forces and decays to zero when the maximum measurable load is applied which varies between models (eg. 100 N). A two-wire interfacing schematic is suggested by the manufacturer: one to supply a reference voltage ($V_{ref}$) (either 5 or 3.3 V) and, one to read the output voltage ($V_{out}$) [26]. A measuring resistor ($R_M$) is also used in the sensor's output circuit to limit the current and to enhance the sensitivity range. The output voltage is computed from Eq. 1.

$$V_{out} = V_{ref} \times \frac{R_M}{R_{FSR} + R_M} \quad (1)$$

The voltage-to-force conversion equation (rarely provided), uses this value to estimate the force. However, since the manufacturer's equation is highly influenced by laboratory test constraints, a more accurate calibration method is proposed to estimate the applied force values from the collected samples.

A digital force gauge measurement instrument (*Sauter* FH-500) with a dynamic measurement range of 0–500 N was used to map the sensors' digital output during sampling process to an accurate numerical force value [29]. The device was mounted on a test stand with an adjustable lever to change the applied forced on the sensors detection area by mechanical displacement of the measurement instrument. Figure 3 shows the calibration interfaces which are used in this phase.

Round-shaped FSR sensors (see Fig. 5) were used in three varieties of size (small, medium, and large) with respect to the mounting position on the human hand. These locations were identified with guidance of the medical experts in research and development stage of our previous experiment. Five sensors were randomly chosen from small and medium size categories for the calibration phase to monitor the sensors' behaviour by changing the applied forces on their detection area. The large sensors were not included in calibration stage since they were only used to indicate the presence or absence of the force exertion on palmar surface of the human hand. Medical students are continuously advised to avoid leaning on the patient's body with this area during palpation procedure to avoid any discomfort. Figure 4 illustrates the thenar and hypothenar eminences on the palmar surface of the human hand.

The sensors' output voltages (0–5 V) were sampled into digital numerical values (0–1023 as 1 byte of data) by on-board analogue-to-digital (ADC block in Fig. 2) converting modules. The actual force magnitudes (in *Newtons*) were simultaneously recorded from the force sensors by a calibration test tool [29] in each force application step. The sensors' digital output was increased by 50 arbitrary units in each calibration step until small changes on the digital output show significant force readings (550 U for small and 750 U for medium sensors).

Despite the noted advantages of the FSR sensors their force measurement reliability is highly dependant on application time. Two undesired behaviours are well described when time is considered in a calibration experiment [30]. The first phenomenon is known as ***Creep*** which is caused by the reduction in sensors' electrical resistance after long-term application of static forces (approximately 2 N higher than the actual force value after 10 min). The second is ***Hysteresis*** reported as the loading and unloading curves (voltage–time plot) were not overlapped. The use of an epoxy resin dome on the sensors' active area is addressed to increase its pressure sensitivity. However, the force measurement range was significantly different when a dome is used particularly when sensors are mounted on the human hand. This may also reduce the human hands' pressure sensitivity and flexibility of the sensors' active area. The loading curve was monitored in this study when participants were asked to reach a given target force; hence, the occurrence of hysteresis was not considered. Also, the total force application duration was identified as 10 s; this helps avoid the behaviour of creep.

**Fig. 3** Calibration interface—*Sauter* FH-500 for actual force value, ParsGlove and DigiScale for digital force value

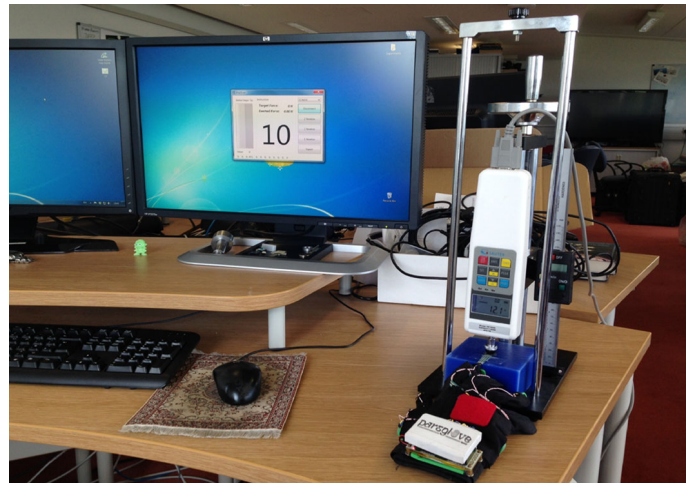

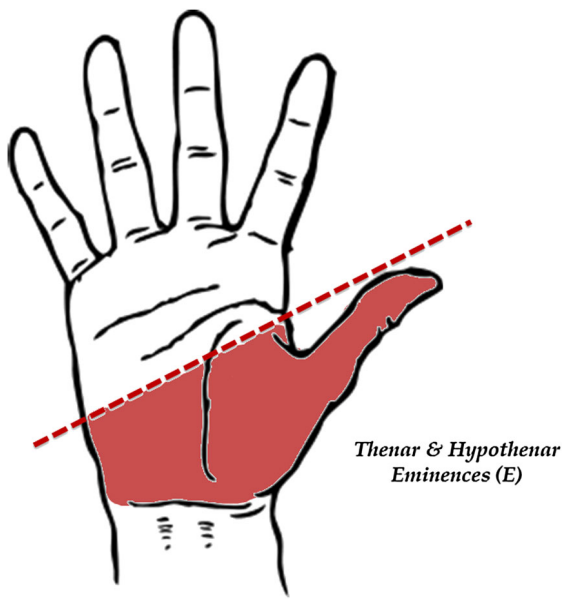

**Fig. 4** Undesirable contact points on palmar surface of the human hand

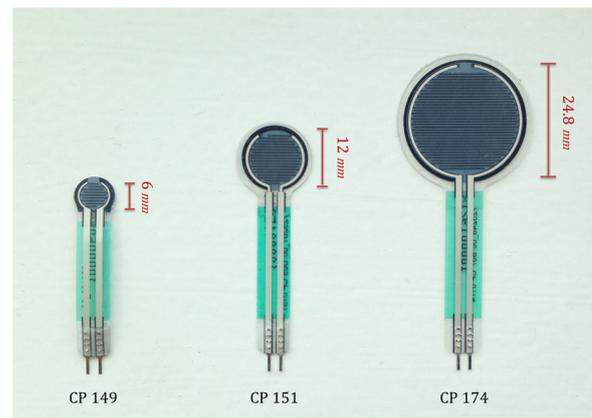

**Fig. 5** IEE *round-shaped* FSR sensors; small (6 mm), medium (12 mm), large (24.8 mm)

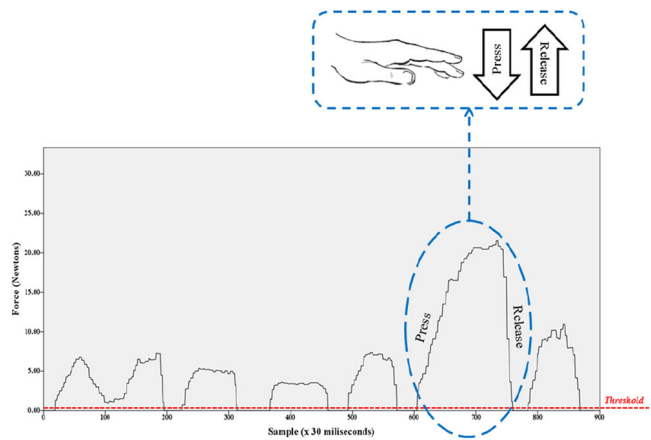

**Fig. 6** Press-release actions during abdominal palpation examination

The force was loaded for a short period of time (approximately 5 s based on our previously captured data from medical experiment) to mimic a hand press-release action (see Fig. 6) and was released from the sensor's detection area soon after each calibration step to avoid appearance of creep and hysteresis effects.

Finally, the sensors' calibration data were plotted (see Fig. 7) to illustrate the best force estimation (red asterisks) in each step and the potential variations among identical sensors (whiskers). Figure 7 shows the calibration outcomes for each category of sensors (small category (6 mm) top, medium category (12 mm) bottom).

A 5th degree polynomial equation (see Eq. 2) was calculated from the calibration data by curve fitting application in MATLAB to accurately estimate the actual forces $F(V)$ applied by the medical users from the sensors' output voltages $V$ that are digitally sampled by the ADC module (0–1023).

$$F(V) = c_5 V^5 + c_4 V^4 + c_3 V^3 + c_2 V^2 + c_1 V + c_0 \quad (2)$$

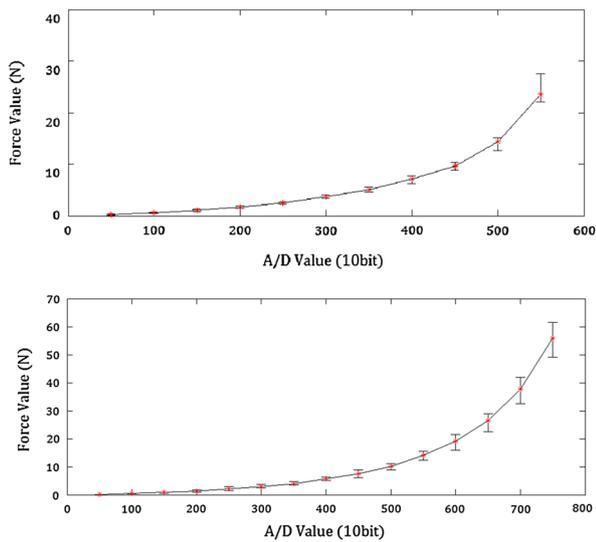

**Fig. 7** Calibration results for five randomly tested force sensors per type

### 3.3 Application

DigiScale is a Graphical User Interface (GUI) specifically developed for this experiment to deliver visual feedback for the force exerted by the tip of users' index fingers. DigiScale has two sections: an information panel providing visual information directly to the user and a toolkit panel for the research investigator. Figure 8 shows a screenshot of DigiScale demonstrating a target force set to 2 N and a sampling timer having collected 3 s of data out of a total of 10 s (with 7 s remaining as shown in the figure).

The information panel is designed to provide visual feedback of applied forces by the user, the target force level in each task and a countdown timer. The toolkit panel is designed to provide connectivity options to pair ParsGlove with DigiScale, three buttons to set the target force levels in each task and finally an export button to persist recorded samples at the end of each session.

The application has capabilities to read continuous data streams over the Bluetooth standard which permits communication with the glove. The only requirement for this application is a Bluetooth transceiver for computers without a built-in module.

### 3.4 Game

To provide game-based training to help improve performance, a 2D game was developed in the Unity 3D game engine. The game assets were adapted from free assets made available from the Unity store. The game was designed to interface with the ParsGlove playing the role of the game controller. Figure 9 illustrates a number of screenshots of the game in action. Figure 10 shows the game as it is being developed.

The goal of the game is to help users improve their pressure sensitivity by controlling a flying bird that soars higher based on the amount of pressure applied by the index finger of the user. The pressure applied is the only form of input. The gameplay is inspired by the infinite runner genre of games; although the play sessions have been limited in the interest of time. The objective of the game is for the bird to collect coins that appear randomly at three possible heights within the environment. The coins are randomly generated in different quantities from 5 to 15. Auditory feedback is also provided on successful coin collection.

The heights chosen correspond to the application of three levels of force. The three force levels are 2, 3 and 5 N, coinciding with very light to medium pressure. These forces were established through a pre-study with medical professionals in a process discussed further in Sect. 4.1. It is crucial for a medical student to control his hand in a dexterous manner to perform different abdominal palpation tasks. Hence, a very light amount of force such as 2 N could be extremely challenging for a novice. Design of different tasks in game-based and application-based approaches were established on these guidelines.

**Fig. 8** The DigiScale application is designed to deliver visual feedback on the exerted force by the user and to collect samples form the user performance

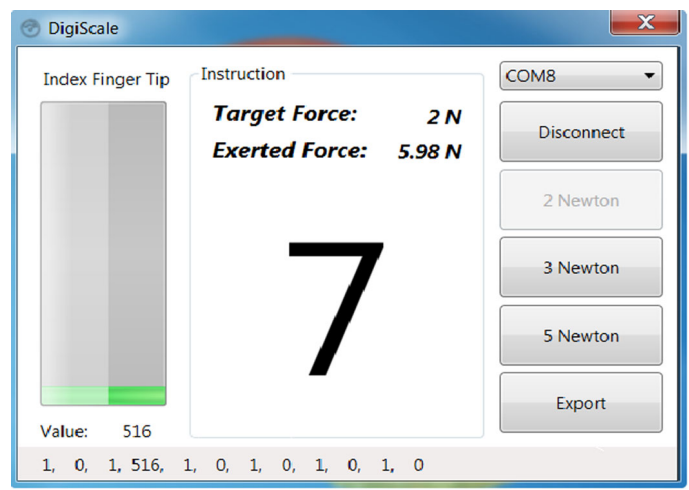

**Fig. 9** A game-based training approach is proposed by this experiment to help users improve their pressure sensitivity by controlling a flying bird that soars higher based on the amount of pressure applied by the index finger of the user. The final figure demonstrates the score that is displayed at the end of a run. Each captured coin is worth 100

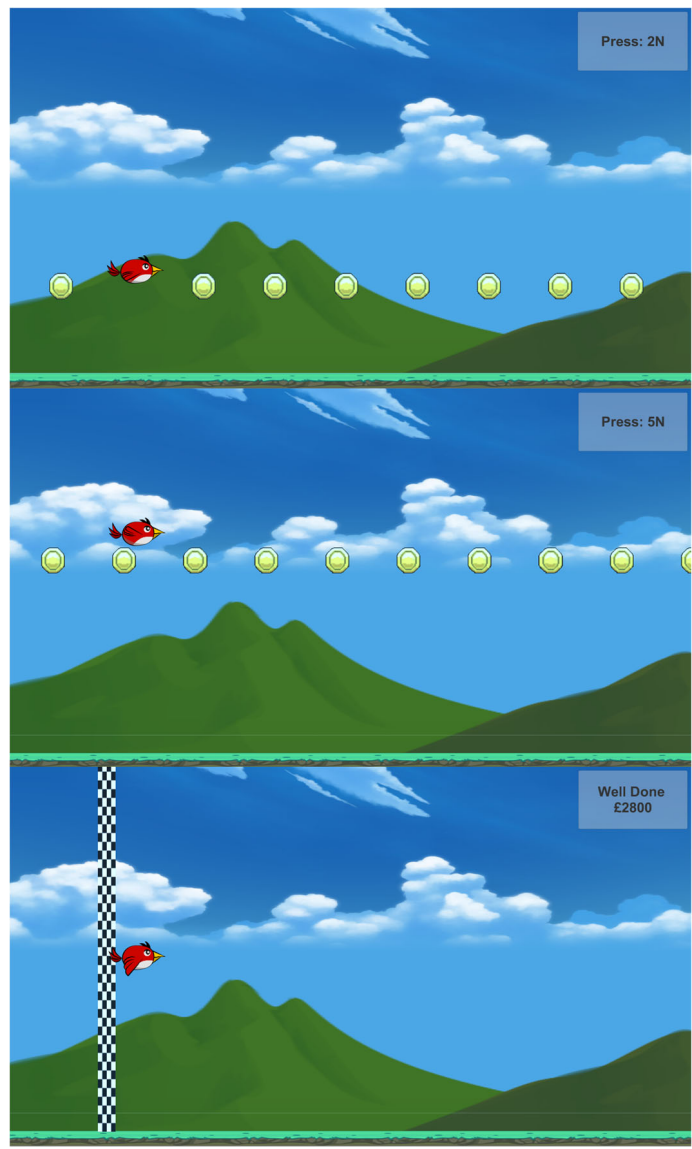

**Fig. 10** The game being developed within the Unity environment

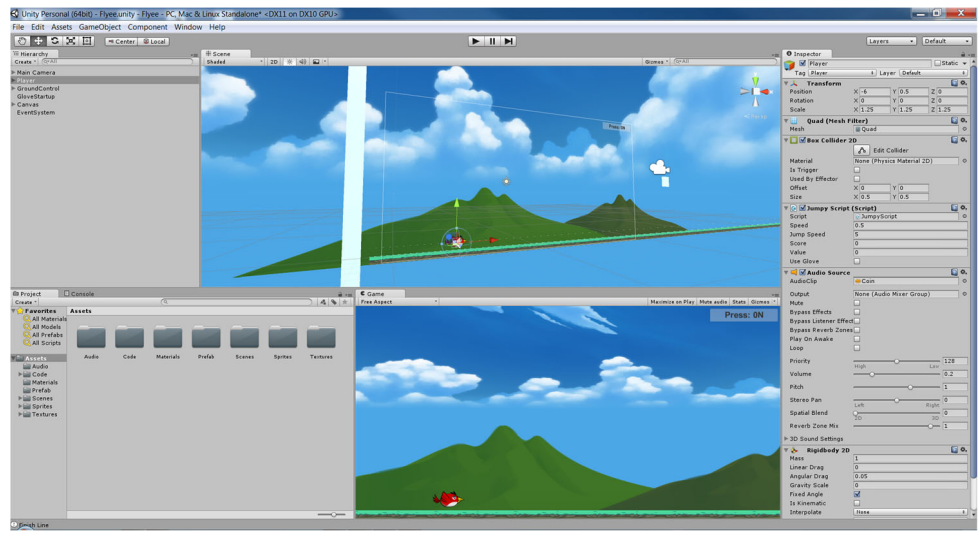

A collision detection function is implemented to detect if the bird avatar hits a coin. The box collider used for coin to bird intersection has a buffer equivalent to ±0.25 N along the height dimension (5 % of the maximum flying altitude 10 N). Players should collect more coins to achieve a better score at the finish line. A slight increase in the flying speed during the game as well as random generation of coins makes it more challenging for the player and has the goal of keeping the game interesting. An information panel is placed on the top right corner of the game screen to show which the next coin level is. This is demonstrated in Newtons for the user to form an association with the amount of pressure to be applied and the numeric value of the force. The information panel also provides a final score at the finish line. The screen height is normalised to represent 0–10 N from bottom to top. Although, the player could reach higher levels of force by pressing harder, the limitation of 10 N was chosen to meet safety regulations.

## 4 Experiment

In order to evaluate whether the game discussed in the previous section improves pressure sensitivity a participant-focused experiment was run. This experiment intends to explore if visual and auditory feedback of applied forces in the form of a game-based training approach could improve motor learning and control abilities on non-medical participants.

### 4.1 Method

A between-participants design was chosen for the experiment. Participants were divided into two groups: Group A ($n = 15$) to be trained by the serious game and group B ($n = 15$) a control group. All participants had been asked to apply force from their index fingertip while sat (in the stand up position kinaesthetic help from shoulder may produce variation in the exertion of force) on a table as rigid surface. The exerted force values were sampled for each target force level for 10 s with 10 ms intervals resulting in 100 samples per each target force level. The three forces of 2, 3, 5 N that were the target goals where the coins were set in the game (as discussed earlier) were based on a pre-study in which medical tutors' pressure while palpating patients was captured. These studies involved the use of four medical professionals examining five different participants acting as patients composed of both genders and three body types. The data capture consisted of all the medical professionals completing three palpation tasks (liver edge, deep and superficial) for all the patients. All data were captured and analysed and the goals of 2, 3, 5 N were identified based on the mean force across the medical professionals in each task across the body types.

**Table 1** Experimental design

|                | Group A              | Group B               |
|----------------|----------------------|-----------------------|
| Training       | Visual feedback (GUI)| Visual feedback (game)|
| Familiarisation| Visual feedback (GUI)| Visual feedback (GUI) |
| Test           | No visual            | No visual             |

Table 1 shows an overview of the experimental design and training methods for each group. The familiarisation phase allowed the participants to acquaint themselves with the equipment and see the actual value they were pressing on DigiScale. In the final test (no visual feedback) the participants could not see how much pressure they were applying on the display and had to rely only on their pressure sensitivity training. The difference between the target value and the recorded value (in Newtons) for the no visual test was used as the dependent variable. The null hypothesis $H0$ in this experiment was that there is no difference between the two groups in the accuracy of the exerted target force for the no visual feedback test session. The software used for the familiarisation DigiScale was significantly different from the environment found in the game to avoid any bias of familiarisation that may have led to the game playing group to have an unfair advantage during the testing phase.

### 4.2 Materials

The primary materials used correspond to the three technologies discussed in Sect. 3. DigiScale was used to convert the raw sensor value from the glove to force in Newtons and to provide visual feedback on the exerted force by the user for the Training and Familiarisation phases. Two TFT displays were used in duplicate mode to provide visual feedback for each participant and to let the research investigator monitor the experiment's progress. An ultra thin powder-coated polyvinyl glove was used to meet hygiene requirements prior to provide the measurement glove to participants. Figure 11 shows a participant from game group in his training session.

### 4.3 Participants

Thirty participants took part in this experiment in two groups of fifteen with seven females and one left-handed participant. Participants all had normal or corrected-to-normal vision. Participants were members of staff or students contacted via internal university email. Participants' age range was between 20 and 55 years with different academic and administrative backgrounds. A participant information leaflet and related ethics documentation were attached to the invitation email before the experiment day to debrief the participants about details prior to the experiment. Participation in this experiment was entirely voluntarily with the right to withdraw at any point.

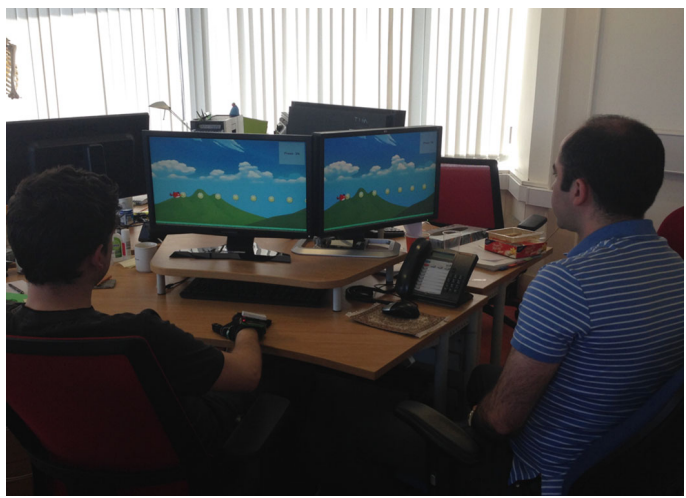

**Fig. 11** A Game-based training session. Synchronisations between visual and kinaesthetic perceptions are key to accurately level the bird's flying altitude

### 4.4 Procedures

Each participant had been debriefed about the experimental steps by research investigators and via email prior to data collection. Each participant was asked to wear a powder-coated ultra thin polyvinyl glove and confirm if the sensor on the index fingertip was positioned correctly.

Group A had 5 min training with DigiScale application. Group B had the same duration of training with the game in three rounds with 1-min. intervals between them. The reason for repeating the training for three attempts for the game group was to provide the player with variety within the game environment as aspects of the game are randomised.

In the familiarisation phase, which occurred soon after the training session, participants were asked to attempt to meet target force levels with the aid of a dedicated display that provided visual feedback via the DigiScale application. In the final no visual feedback test, the display was switched off and results collected for each participant. There was a 1-min. interval between training and tests to avoid human fatigue. For each target force the following objectives need to be achieved by each participant in each test:

- To reach the given target force level
- To maintain that target force level for 10 s.

### 4.5 Results

Results for each target force was obtained via the difference in target and recorded force for each observation. The mean exerted force ($\mu_i$) for each target force level ($f_i$) is calculated from collected samples for each participant. The absolute difference from the target force is calculated as:

$$\delta_i = |\mu_i - f_i|$$

The mean of the delta values for all three target force levels $f_i$ (2, 3 and 5 N) were computed as a final result for each participant. A non-parametric Mann–Whitney test has been selected to analyse the results due to the non-parametric nature of the data.

The accuracy in the exerted target force for the no-visual test for participants in group B, who were trained using our game approach (Mdn = 0.86), differed significantly from the participants in group A, who trained using the application only approach (Mdn = 1.56), $U = 61$, $z = -2.137$, $p < 0.05$, $r = -0.36$ and thus $H0$ is rejected.

A post-hoc power analysis test is power analysis was conducted, using the *G*power* software package [31], to determine the likelihood of detecting the true effect size. The statistical power $(1 - \beta)$ is function of the type I error ($\alpha = .05$), the size of the measured effect ($d = .85$), and the total number of participants in the study ($N = 30$). Power calculation result shows three extra participant per group ($n = 18$) could improve the obtained power (0.74) to the recommended level (0.80) [32].

This result may highlight the potential role of game-based training on cognitive and control motor learning abilities. One possible reason for this achievement is an improvement in the understanding of the approximate force and sensitivity for the required pressure instilled while playing the game. Another potential advantage of game-based training is the competition factor characteristic of games. It was observed during the experiment that participants in group B were keen to beat their previous best score in each round which may have led to better focus and concentration on the requested test.

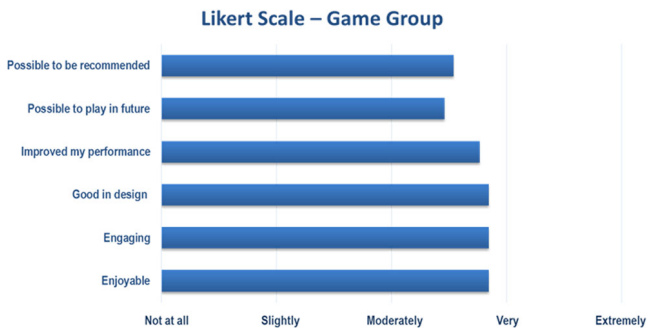

**Fig. 12** Feedback from the game group

### 4.6 Qualitative feedback

In order to form an understanding of whether the game was considered an enjoyable experience and whether it was well designed and engaging a number of questions were asked to the group that played the game. An electronic questionnaire was sent via email to each participant in the game group ($n = 15$) to collect their reflective feedback on their experience when playing the game. A total of six questions were asked to rank key features of the game from 1 to 5 (e.g. for first question the answer is made from Not at all, Slightly, Moderately, Very, Extremely "Enjoyable"). Fourteen out of fifteen participants replied. Figure 12 illustrates the feedback in a Likert scale.

Using rounded mean scores for a general evaluation the game can be deemed to be very enjoyable to play, very engaging, well designed, and with the ability to provide a perceived increase in motor ability. Participants also considered that they were likely to play the game again given the opportunity and would very much recommend it to a friend. On the whole, based on this feedback, the game design seems to have been for the most part successful.

## 5 Conclusions and future work

This work has presented a serious game that attempts to teach participants the correct application of pressure by controlling a virtual character on screen via a pressure sensitive input device that rewards players with accurate and controlled input. ParsGlove, an input device for measuring pressure sensitivity, has also been introduced as the main input device for playing and training. The results demonstrate that those players that played the game performed significantly better than a control group in a subsequent no-visual task within a very different environment from the game itself. Moreover, questionnaire responses indicate that the game is enjoyable and engaging. While this game has appeared to have been successful, it is part of a larger framework that is required to make automated or assisted palpation training successful [9].

Future work will look into enhancing this experience using all the input sensors on the glove and the capability of the system to capture location and orientation data. This can be then used for the development of a serious simulator or serious game. Furthermore, palpation is not the only application that requires pressure sensitivity and modifications to the main game to adapt to the range of sensitivities of various applications can aid pressure sensitivity training in other fields equation training for musical instruments.

**Acknowledgements** Grateful acknowledgement is made to members of staff and students in WMG for their participation and support. We would like to thank Unity and Creepy Cat for making available the 2D Game Starter Assets used to prepare the game used in this publication. Debattista is partially supported by a Royal Society Industrial Fellowship.

## References

1. Chalmers, A., Debattista, K., Ramic-Brkic, B.: Towards high-fidelity multi-sensory virtual environments. Vis. Comput. **25**(12), 1101–1108 (2009)
2. Arnab, S., Dunwell, I., Debattista, K.: Serious Games for Healthcare: Applications and Implications: Applications and Implications. IGI Global, Pennsylvania (2012)
3. Sliney, A., Murphy, D.: Jdoc: a serious game for medical learning. In: 2008 First International Conference on Advances in Computer–Human Interaction, pp. 131–136 (2008)
4. Susi, T., Johannesson, M., Backlund, P.: Serious Games: An Overview (2007)
5. Macleod, J., Douglas, G., Nicol, E.F., Robertson, C.E.: Macleod's Clinical Examination. Elsevier Health Sciences, Amsterdam (2009)
6. Patel, V., Morrissey, J.: Practical and Professional Clinical Skills. Oxford University Press, Oxford (2011)
7. Goodwin, J.: The importance of clinical skills. BMJ **310**(6990), 1281–1282 (1995)
8. Hulusic, V., Harvey, C., Debattista, K., Tsingos, N., Walker, S., Howard, D., Chalmers, A.: Acoustic rendering and auditory–visual cross-modal perception and interaction. In: Computer Graphics Forum, vol. 31, no. 1, pp. 102–131. Wiley Online Library, New York (2012)
9. Asadipour, A.: A technology-aided multi-modal training approach to assist abdominal palpation training and its assessment in medical education. Ph.D. Dissertation, University of Warwick (2005)

10. Duvivier, R.J., van Geel, K., van Dalen, J., Scherpbier, A.J., van der Vleuten, C.P.: Learning physical examination skills outside timetabled training sessions: what happens and why? Adv. Health Sci. Educ. **17**(3), 339–355 (2012)
11. Bendtsen, L., Jensen, R., Jensen, N., Olesen, J.: Pressure-controlled palpation: a new technique which increases the reliability of manual palpation. Cephalalgia **15**(3), 205–210 (1995)
12. Futarmal, S., Kothari, M., Ayesh, E., Baad-Hansen, L., Svensson, P.: New palpometer with implications for assessment of deep pain sensitivity. J. Dental Res. **90**(7), 918–922 (2011)
13. Burdea, G., Patounakis, G., Popescu, V., Weiss, R.E.: Virtual reality-based training for the diagnosis of prostate cancer. IEEE Trans Biomed Eng **46**(10), 1253–1260 (1999)
14. Gotsis, M.: Games, virtual reality, and the pursuit of happiness. Comput. Graph. Appl. IEEE **29**(5), 14–19 (2009)
15. Scarle, S., Dunwell, I., Bashford-Rogers, T., Selmanovic, E., Debattista, K., Chalmers, A., Powell, J., Robertson, W.: Complete motion control of a serious game against obesity in children. In: 2011 third international conference on games and virtual worlds for serious applications (VS-GAMES), pp 178–179 (2011)
16. Schonauer, C., Pintaric, T., Kaufmann, H., Jansen Kosterink, S., Vollenbroek-Hutten, M.: Chronic pain rehabilitation with a serious game using multimodal input. In: 2011 international conference on virtual rehabilitation (ICVR), pp 1–8 (2011)
17. Saini, S., Rambli, D., Sulaiman, S., Zakaria, M., Shukri, S.: A low-cost game framework for a home-based stroke rehabilitation system. In: 2012 international conference on computer information science (ICCIS), vol 1, pp. 55–60 (2012)
18. Kamel Boulos, M.N.: Healthcybermap (2011). http://www.healthcybermap.org/. **(Online)**
19. Graafland, M., Schraagen, J.M., Schijven, M.P.: Systematic review of serious games for medical education and surgical skills training. Brit. J. Surg. **99**(10), 1322–1330 (2012). **(Online)**. doi:10.1002/bjs.8819
20. Dunwell, I., Jarvis, S.: A serious game for on-the-ward infection control awareness training: ward off infection. In: Serious Games for Healthcare: Applications and Implications, pp. 233 (2012)
21. Kato, P.M., Cole, S.W., Bradlyn, A.S., Pollock, B.H.: A video game improves behavioral outcomes in adolescents and young adults with cancer: a randomized trial. Pediatrics **122**(2), e305–e317 (2008)
22. Carmeli, E., Vatine, J.-J., Peleg, S., Bartur, G., Elbo, G.: Upper limb rehabilitation using augmented feedback: impairment focused augmented feedback with handtutor. In: Virtual rehabilitation international conference, pp. 220–220 (2009)
23. Kamel Boulos, M.N., Hetherington, L., Wheeler, S.: Second life: an overview of the potential of 3-d virtual worlds in medical and health education. Health Inf. Librar. J. **24**(4), 233–245 (2007). doi:10.1111/j.1471-1842.2007.00733.x. **(Online)**
24. Brown, K.E., Bayley, J., Newby, K.: Serious game for relationships and sex education: application of an intervention. In: Serious Games for Healthcare: Applications and Implications: Applications and Implications, p. 135 (2012)
25. Srinivasan, M.A., Chen, J.-S.: Human performance in controlling normal forces of contact with rigid objects. ASME Dyn Syst Control Div Publ DSC, vol. 49, pp. 119–125. ASME, New York (1993)
26. IEE (2013) Customized input sensing—CIS solutions (2013). http://www.iee.lu/. **(Online)**
27. Jensen, T.R., Radwin, R.G., Webster, J.G.: A conductive polymer sensor for measuring external finger forces. J. Biomech. **24**(9), 851–858 (1991)
28. Williams, E.M., Gordon, A.D., Richmond, B.G.: Hand pressure distribution during oldowan stone tool production. J. Human Evol. **62**(4), 520–532 (2012)
29. Sauter, GmbH: Instruction manual for force gauge fh series (2015). http://www.force-gauges.co.uk/pdf/fh-m.pdf. **(Online)**
30. Florez, J. et al.: Calibration of force sensing resistors (fsr) for static and dynamic applications. In: ANDESCON, 2010 IEEE, pp. 1–6. IEEE, New York (2010)
31. Faul, F., Erdfelder, E., Lang, A.-G., Buchner, A.: G* power 3: a flexible statistical power analysis program for the social, behavioral, and biomedical sciences. Behav. Res. Meth. **39**(2), 175–191 (2007)
32. Cohen, J.: Statistical Power Analysis for the Behavioral Sciences. Lawrence Erlbaum Associates Inc, NJ (1977)

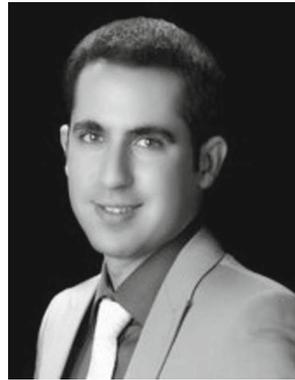

**Ali Asadipour** holds an engineering PhD, and a MSc in computer science from University of Warwick. His research interests are in multi-modal virtual and augmented reality simulations, human factors and ergonomics and computer informatics

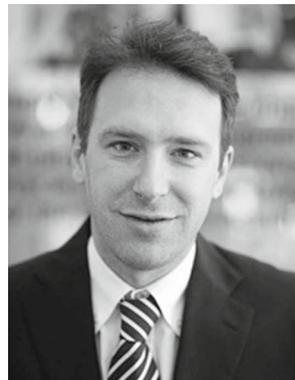

**Kurt Debattista** holds a PhD from the University of Bristol, an MSc in Computer Science, an MSc in Psychology and a BSc in Mathematics. Dr Debattista has published over 70 papers in peer-reviewed international conferences and journals, four guest editorials in leading journals, co-edited six books and coauthored one book on advanced high dynamic range imagery. Dr Debattista is a member of a number of programme committees for international conferences. In 2013, he was also granted a Royal Society Industrial Fellowship in co-operation with Jaguar Land Rover.

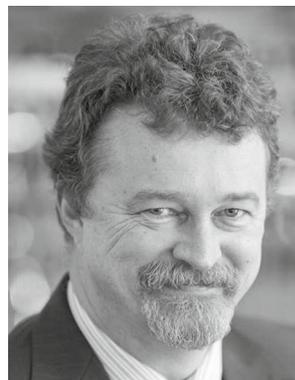

**Alan Chalmers** is a Professor of Visualisation at WMG, University of Warwick, UK and a Royal Society Industrial Fellow. He has an MSc with distinction from Rhodes University, 1985 and a PhD from University of Bristol, 1991. He is Honorary President of Afrigraph and a former Vice President of ACM SIGGRAPH. Chalmers has published over 220 papers in journals and international conferences on HDR, high-fidelity virtual environments, multi-sensory perception, parallel processing and virtual archaeology and successfully supervised 37 PhD students.